# Does Embodiment Matter to Biomechanics and Function? A Comparative Analysis of Head-Mounted and Hand-Held Assistive Devices for Individuals with Blindness and Low Vision

**Authors**:
Gaurav Seth [a], Hoa Pham [b], Giles Hamilton-Fletcher [c], Charles Leclercq [d], John-Ross Rizzo [b, c, e, f]

**Affiliation**:
[a] Steinhardt School of Culture, Education, and Human Development, New York University, New York, NY, USA
[b] Rusk Rehabilitation, NYU Langone Health, New York, NY, USA
[c] Department of Rehabilitation Medicine, NYU Grossman School of Medicine, New York, NY, USA
[d] ARxVision LLC, London, UK and New York, NY, USA
[e] Department of Ophthalmology, NYU Grossman School of Medicine, New York, NY, USA
[f] Department of Neurology, NYU Langone Health, New York, NY, USA

**Contact**:
Gaurav Seth
gs3823@nyu.edu

## ABSTRACT

Visual assistive technologies, such as Microsoft's Seeing AI, can improve access to environmental information for persons with blindness or low vision (pBLV). Yet, the physical and functional implications of different device embodiments remain unclear. In this study, 11 pBLV participants used Seeing AI on a hand-held smartphone and on a head-mounted ARx Vision system to perform six activities of daily living, while their movements were captured with Xsens motion capture. Functional outcomes included task time, success rate, and number of attempts, and biomechanical measures included joint range of motion, angular path length, working volume, and movement smoothness. The head-mounted system generally reduced upper-body movement and task time, especially for document-scanning style tasks, whereas the hand-held system yielded higher success rates for tasks involving small or curved text. These findings indicate that both embodiments are viable, but they differ in terms of physical demands and ease of use. Incorporating biomechanical measures into assistive technology evaluations can inform designs that optimise user experience by balancing functional efficiency, physical sustainability, and intuitive interaction.

**INTRODUCTION**

Based on recent estimates from the World Health Organization, there are at least 2.2 billion persons with blindness or low vision (pBLV) globally. Visual impairments can impede many fundamental daily tasks, including recognizing text, fully understanding environments, and avoiding obstacles during navigation. The ability to parse text and identify objects in one's immediate environment is essential to routine medication management, community and household navigation, meal preparation, and other activities contributing to functional independence, otherwise known as activities of daily living (ADL) (Gamage et al. 2023; Golubova et al. 2021).

Assistive technologies (AT) are designed to support pBLV in completing ADLs more safely and effectively. Traditional ATs include tools like white canes and Braille-based devices. More recently, smartphone-based applications leveraging artificial intelligence (AI) and computer vision have emerged as powerful aids for functional problems arising from visual impairments (Han et al. 2024; Golubova et al. 2021; Plikynas et al. 2020; Hanlu et al. 2014; Neat et al. 2019). One widely used app is Seeing AI by Microsoft, which can read text, recognize faces, describe scenes, and identify currency and products ([seeingAI.com](seeingAI.com)). Smartphones serve as convenient hosts for such tools, offering portability and processing power in one compact system.

The hand-held nature of smartphone-based ATs requires continuous user engagement, introducing ergonomic challenges not present in hands-free, body-mounted alternatives. These concerns are supported by evidence linking smartphone overuse to musculoskeletal disorders, including neck pain, hand dysfunction, and upper limb strain (Eitivipart, Viriyarojanakul, and Redhead 2018; Barrett, McKinnon, and Callaghan 2020; İnal et al. 2015; Soliman Elserty, Ahmed Helmy, and Mohmed Mounir 2020). Such risks are especially relevant for pBLV, who already experience higher rates of musculoskeletal discomfort (Kolli et al. 2022), often associated with the prolonged and repetitive physical demands of mobility aids like canes and guide dogs (Julie Mount,

Gitlin, and Howard 1997; McCall et al. 2021; Gitlin et al. 1997). These demands have been linked to repetitive wrist and arm movements and reduced intersegmental coordination (J. Mount et al. 2001). Additional strain from smartphone use, for example, taking multiple photographs from different angles, may further exacerbate fatigue and potentially contribute to overuse injuries, particularly in the neck, wrist, and upper limbs. Compensatory neck and arm movements used to orient in space or operate assistive technologies may also increase cumulative physical burden, particularly during extended periods of device use. Considering these factors, reducing musculoskeletal strain through ergonomic redesign or behavioral counseling may help mitigate the physical burden associated with smartphone overuse in this population (Soliman Elserty, Ahmed Helmy, and Mohmed Mounir 2020; Han et al. 2024).

Beyond physical ergonomics, several social and technological constraints also influence the usability of smartphone-based ATs. For instance, aiming a smartphone camera is neither discreet nor effortless and raises significant concerns related to user privacy and situational awareness, since users may not notice people around them while trying to aim the device (Ahmed et al. 2017; Kuribayashi et al. 2021). Additionally, the need for real-time connectivity to access cloud-based AI services introduces practical constraints, such as battery life, network dependence, and latency from remote data processing (Azzino et al. 2024; Yuan et al. 2022; Hamilton-Fletcher et al. 2024).

Given these concerns, hands-free wearable devices such as head-mounted smart glasses may offer an ergonomic advantage. Smart glasses align with natural head and gaze direction, reducing the need for additional arm movements. They are also less conspicuous, improving safety and privacy in public settings (Kuribayashi et al. 2021; Gamage et al. 2023). In user surveys, pBLV have reported a preference for smart glasses over smartphones for reasons including hands-free operation, natural gaze alignment, and improved usability for specific populations like guide-dog users and musicians (Gamage et al. 2023; Lussier-Dalpé et al. 2022; Hanlu et al. 2014).

Despite these reported advantages, few studies have quantitatively compared the biomechanical demands and functional performance of smartphones versus head-mounted devices during real-world tasks. Understanding these differences is important, as repeated upper body strain may lead not only to discomfort but also to long-term disability in a population already vulnerable to musculoskeletal risk (Kolli et al. 2022). Moreover, effective and efficient environmental-interaction tools are essential to promoting independence in pBLV (Golubova et al. 2021; Plikynas et al. 2020).

This study addresses this gap by evaluating the biomechanical and functional impact of using a head-mounted device (ARx Vision) versus a smartphone (Google Pixel 4) both running Seeing AI. Upper-body biomechanics were recorded using the Xsens motion capture system, capturing both angular and translational kinematics of the head, trunk, and hands. Key biomechanical parameters analyzed included joint range of motion (ROM), angular path length, working volume, and movement smoothness. In addition, functional performance outcomes—such as task completion time, success rate, and number of attempts—were evaluated. By comparing these measures across the two assistive-technology embodiments, the study aim was to identify which configuration more effectively supports safe, efficient, and successful performance of ADLs among pBLV.

## METHODS

### Experimental Protocol

There were eleven participants in the study (7 females and 4 males) with varying degrees of visual impairment, as detailed in Table 1. The study protocol was reviewed and approved by the Institutional Review Board (IRB) at the Lighthouse Guild to ensure ethical compliance, and informed consent was obtained from all participants after explaining the purpose, procedures, and potential risks of the study.

Table 1: Participant demographics and binocular visual acuity reported using logMAR (logarithm of the minimum angle of resolution) values. A logMAR of 0.0 corresponds to normal vision (20/20 Snellen equivalent), with higher values indicating worse visual acuity.

| Participant | Age | Gender | Cause of Visual Impairment | LogMAR |
|---|---|---|---|---|
| 1 | 37 | Female | Uveitis & Band Keratopathy | 1.40 |
| 2 | 71 | Male | Congenital Optic Nerve Atrophy | 0.82 |
| 3 | 49 | Female | Retinitis Pigmentosa | 1.40 |
| 4 | 57 | Male | Glaucoma | 2.00 |
| 5 | 64 | Male | Leber Congenital Amaurosis | 1.40 |
| 6 | 63 | Female | Uveitis | 0.66 |
| 7 | 66 | Female | Acute Congenital Glaucoma | 2.00 |
| 8 | 54 | Female | Stargardt Disease | 1.34 |
| 9 | 83 | Male | Macular Degeneration | 0.82 |

| 10 | 60 | Female | Cone-rod Dystrophy | 1.40 |
| 11 | 61 | Female | Stargardt Disease | 1.10 |

**Instrumentation**

The study utilized two assistive-technology configurations. The head-mounted setup featured the ARx Vision Gen 1.5, which includes a bone conduction audio headset and a camera module, used in conjunction with a Google Pixel 4 smartphone running Android 11. The hand-held configuration used the same model smartphone independently. Both setups ran Seeing AI version 0.7.14. To capture upper-body biomechanics, the Xsens MVN Awinda system was employed, with 11 sensors placed on key anatomical landmarks including the head, pelvis, sternum, and bilateral hands, forearms, upper arms, and shoulders. The instrumentation schematics are shown in figure 1.

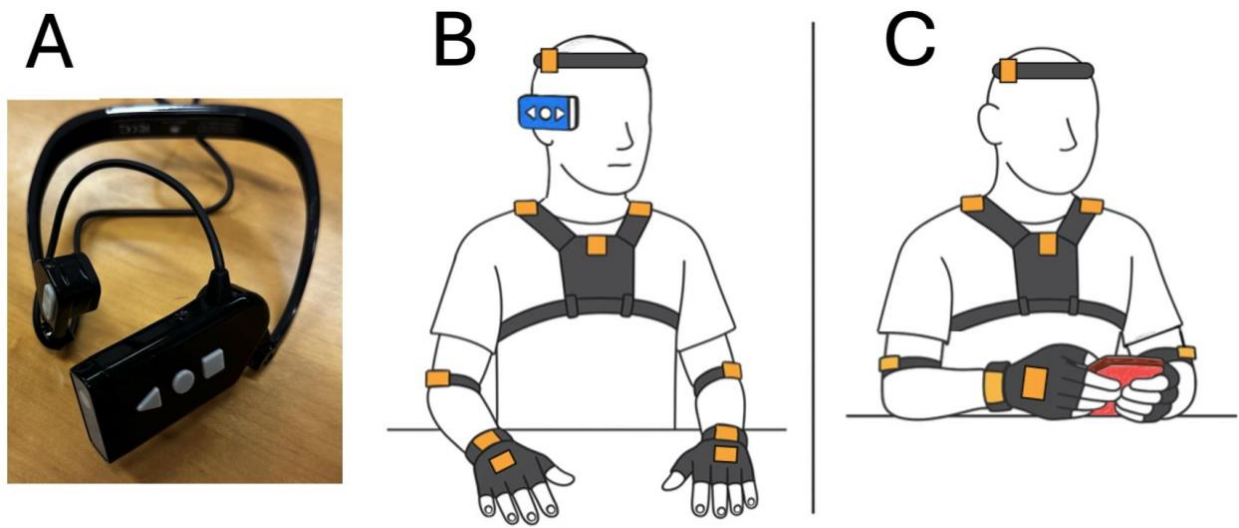

*Figure 1: Instrumentation used in the study. (A) Head-mounted ARx Vision Gen 1.5 with bone conduction headset and camera module, paired with a Google Pixel 4 smartphone running Seeing AI. (B) Head-mounted configuration and (C) Hand-held smartphone configuration, both shown with motion capture instrumentation.*

**Task Overview**

Each participant performed six tasks using the Seeing AI application across two embodiments: hand-held smartphone or a head-mounted ARx Vision device. The order of embodiment was counterbalanced to avoid order effects. Tasks utilized specific Seeing AI modes (e.g., short text, document, person), and participants received training on relevant modes prior to starting the protocol. Instructions and follow-up questions for each task were standardized for consistency. Before each task, participants were instructed to establish a baseline hand position to standardize the starting point (e.g., for seated tasks, hands were placed flat on the table). Participants were permitted multiple attempts to complete each task provided they followed the predefined criteria.

**Task Descriptions**

The study involved six functional tasks simulating common activities of daily living for individuals with visual impairments. Tasks were performed in a fixed order under both device embodiments, as listed below:

1. Article Reading (Document Mode): Participants read and summarized a newspaper article.
2. Invoice Reading (Document Mode): Participants determined the total amount due from an invoice.
3. Medicine Identification (Short Text Mode): Participants identified the name and dosage instructions on a medicine bottle.
4. Product Identification (Short Text Mode): Participants located a can of specified product from a shelf with four different cans.
5. Person Detection (Person Mode): Participants identified the location of a person in a room.
6. Street Sign Identification (Short Text Mode): Participants read a street name on a sign positioned 8 meters away.

For brevity, we refer to these tasks throughout the manuscript as the Article, Invoice, Medicine, Product, Person, and Street Sign tasks, respectively.

**Data Analysis**

The data were analyzed across two main domains: functional performance and biomechanical measures, each capturing different aspects of how participants performed the tasks.

To compare the head-mounted and hand-held systems, the primary analysis focused on the final attempt of each task, reflecting real-world usage where participants typically stopped once they obtained a satisfactory response. A secondary analysis examined performance across all attempts to explore summative effects and broader usage patterns. In this secondary analysis, task time was summed across all attempts; joint ROM was taken as the maximum value across attempts; while angular path length and working volume were both summed across attempts; and summative movement smoothness (LDLJ-V) was calculated by summing jerk cost across all attempts before applying a logarithmic transformation and normalization to yield a single dimensionless value.

To enable comparisons across tasks, global scores were calculated by averaging each biomechanical parameter across all six tasks for each participant. This yielded one representative value per participant per body segment, separately for each system.

For statistical calculations, outliers were excluded if they exceeded three standard deviations ($\mu \pm 3\sigma$) from the mean. Due to the limited sample size, normality could not be reliably assessed; therefore, non-parametric methods were used to avoid assumptions about data distribution. System-level comparisons were made using the Wilcoxon signed-rank test. To control the false discovery rate during multiple comparisons, Benjamini-Hochberg (BH) correction was applied (Benjamini and Hochberg 1995). Statistical significance was assessed using BH-corrected *p*-values, with $p < 0.05$ indicating significance for applicable measures. Observational trends are reported for measures with count variables where no statistical tests were conducted.

The following section provides definitions and calculation methods for each parameter. Before estimating any functional or biomechanical parameter, data was trimmed by identifying the start and end points of movement using signals from the initiating body segment. A one-second baseline of static activity was used to compute the mean ($\mu$) and standard deviation ($\sigma$), with task initiation defined as the point where the signal exceeded $\mu \pm 2.5\sigma$. For non-verbal tasks, completion was detected using a similar threshold, whereas, for verbal tasks, the recording was manually stopped after the response. All task initiation and completion points were plotted and manually verified before proceeding with further analyses.

**Functional Performance Analysis**

Functional measures evaluated task performance outcomes and included the following metrics:

***Task Time:*** Task duration was calculated as the total time taken for task completion in the trimmed data set.

***Percentage of Successful Participants:*** This metric represents the proportion of participants who successfully completed each task, regardless of the number of attempts made. The percentage was calculated by dividing the number of participants with at least one successful completion by the total number of participants. This was computed separately for each task and is reported descriptively, without inferential statistical analysis.

***Number of Attempts:*** The total count of retries or repeated attempts required by participants to complete a task, as another descriptive measure.

**Biomechanical Data Analysis**

All biomechanical measures were calculated using a temporally trimmed dataset, focusing exclusively on the duration during which participants performed the task. This trimming excluded any time periods before and after task execution.

***Working Volume:*** The working volume was calculated to quantify the spatial volume covered by participants during task execution for the right-hand, left hand, and head segments. Positional data (x, y, z coordinates) were extracted between the identified start and end Indices. A three-dimensional convex hull was constructed to define the movement boundary, and the enclosed volume (V) was computed using a triangulation method, summing the volumes of individual tetrahedra. This process was conducted separately for each body segment. The working volume for a representative participant and task is illustrated in Figure 2.

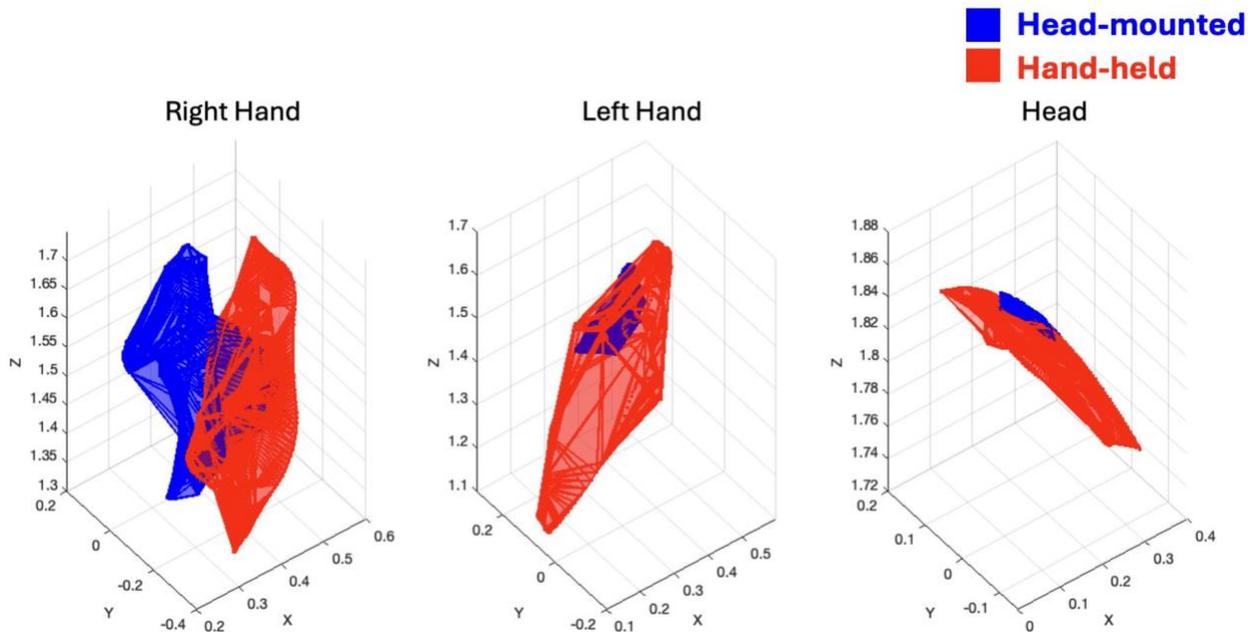

*Figure 2: Representative working volume for the right hand, left hand, and head during invoice task execution for a single participant. The 3D convex hull for the hand-held system is depicted in red, while the head-mounted system is shown in blue, illustrating the spatial boundaries covered by each system during movement.*

***Joint Range of Motion (ROM):*** Joint ROM quantified angular excursions of the head, trunk (T12), right hand, and left hand during task execution. Joint angle data (θx, θy, θz) were extracted from the Start to End Index. ROM was computed as the difference between maximum and minimum angles for each axis (x, y, z) and combined using the Euclidean norm:

$$ROM = \sqrt{(\Delta\theta_x)^2 + (\Delta\theta_y)^2 + (\Delta\theta_z)^2}$$

This process was repeated independently for each body segment, yielding four ROM values per task attempt.

***Angular Path Length:*** Angular path length quantifies the total angular motion of each body segment during task execution. This metric was calculated for the head, trunk (T12), right hand, and left hand. Joint angle data (θx, θy, θz) were extracted from the trimmed time window and converted to a scalar angle magnitude at each time point using the Euclidean norm:

$$Angle\ Magnitude = \sqrt{(\theta_x)^2 + (\theta_y)^2 + (\theta_z)^2}$$

The angular path length was then computed as the sum of absolute differences between consecutive angle magnitudes, representing the cumulative angular motion of the body segment over time. This yielded a single value per body segment per task attempt.

***Movement Smoothness:*** LDLJ-V was computed to assess movement smoothness for the right hand, left hand, and head during task execution, following the methodology of Melendez-Calderon et al. (Melendez-Calderon, Shirota, and Balasubramanian 2020). Velocity data ($v_x, v_y, v_z$) were extracted from task onset to completion. Jerk, the rate of change of acceleration, was calculated as the second derivative of velocity

$$\text{Jerk} = \frac{d^2 Velocity}{dt^2}$$

where dt = 1/60 seconds (sampling rate: 60 Hz). The jerk magnitude was computed as:

$$Jerk\ Magnitude = \sqrt{(Jerk_x)^2 + (Jerk_y)^2 + (Jerk_z)^2}$$

The cumulative jerk cost was obtained via numerical integration (trapezoidal rule), then normalized using task duration (T) and the peak velocity ($V_{peak}$)

$$\text{Normalized Jerk} = \frac{T^3}{V_{peak}^2} \times \int (\text{Jerk Magnitude})^2\ dt$$

Finally, the LDLJ-V was computed as the negative logarithm of the normalized jerk:

$$\text{LDLJ-V} = -\log(\text{Normalized Jerk})$$

This calculation was performed independently for the right hand, left hand, and head.

**RESULTS**

**Functional Performance Measures**

*Task Time:* Task completion time was generally faster with the head-mounted (HM) as compared to hand-held (HH) system, with significant reductions observed in the Article reading and Person tasks. This trend was reflected in the global task time, which was also significantly lower for the head-mounted system (Figure 3A, Table 2).

*Percentage of Successful Participants:* The percentage of successful participants was similar across systems for most tasks. However, the hand-held system showed higher success rates in the Article, Invoice, and Medicine Identification tasks, contributing to a higher global completion rate overall (HM: 86.37%, HH: 95.46%) (Figure 3B, Table 2).

*Number of Attempts:* For the number of attempts, the head-mounted system required more attempts on average, especially in the Invoice, Product, and Article tasks. The global average number of attempts was higher with the head-mounted system (HM: 1.4 vs. HH: 1.2), although no inferential statistical tests were applied to this metric (Figure 3C, Table 2).

*Table 2: **Functional Performance Metrics Across Tasks and Embodiments for Final Attempt**. The table summarizes the average task performance metrics across all participants for each task, comparing the head-mounted (HM) and hand-held (HH) systems. Metrics include task time (in seconds), percentage of successful participants, and number of attempts. The third row under task time reports Benjamini-Hochberg (BH) corrected p-values from Wilcoxon signed-rank tests. Shaded cells indicate statistically significant differences ($p < 0.05$).*

| Embodiment Type | Article | Invoice | Medicine | Product | Street Sign | Person | Global Score |
|---|---|---|---|---|---|---|---|
| **Task Time (seconds)** | | | | | | | |
| HM | 114.0 | 60.2 | 69.0 | 74.0 | 37.3 | 10.1 | **60.8** |
| HH | 202.2 | 71.4 | 77.4 | 42.8 | 24.4 | 15.1 | **72.2** |
| *p* | *0.006* | *0.496* | *0.638* | *0.107* | *0.152* | *0.006* | ***0.024*** |
| **Percentage of Successful Participants (%)** | | | | | | | |
| HM  | 90.91 | 90.91 | 54.55 | 100 | 81.82 | 100 | **86.37** |
| HH  | 100   | 100   | 90.91 | 100 | 81.82 | 100 | **95.46** |
| **Number of Attempts** | | | | | | | |
| HM  | 1.6   | 1.6   | 1.4   | 1.5 | 1.1   | 1.1 | **1.4** |
| HH  | 1.4   | 1.2   | 1.3   | 1.1 | 1.1   | 1   | **1.2** |

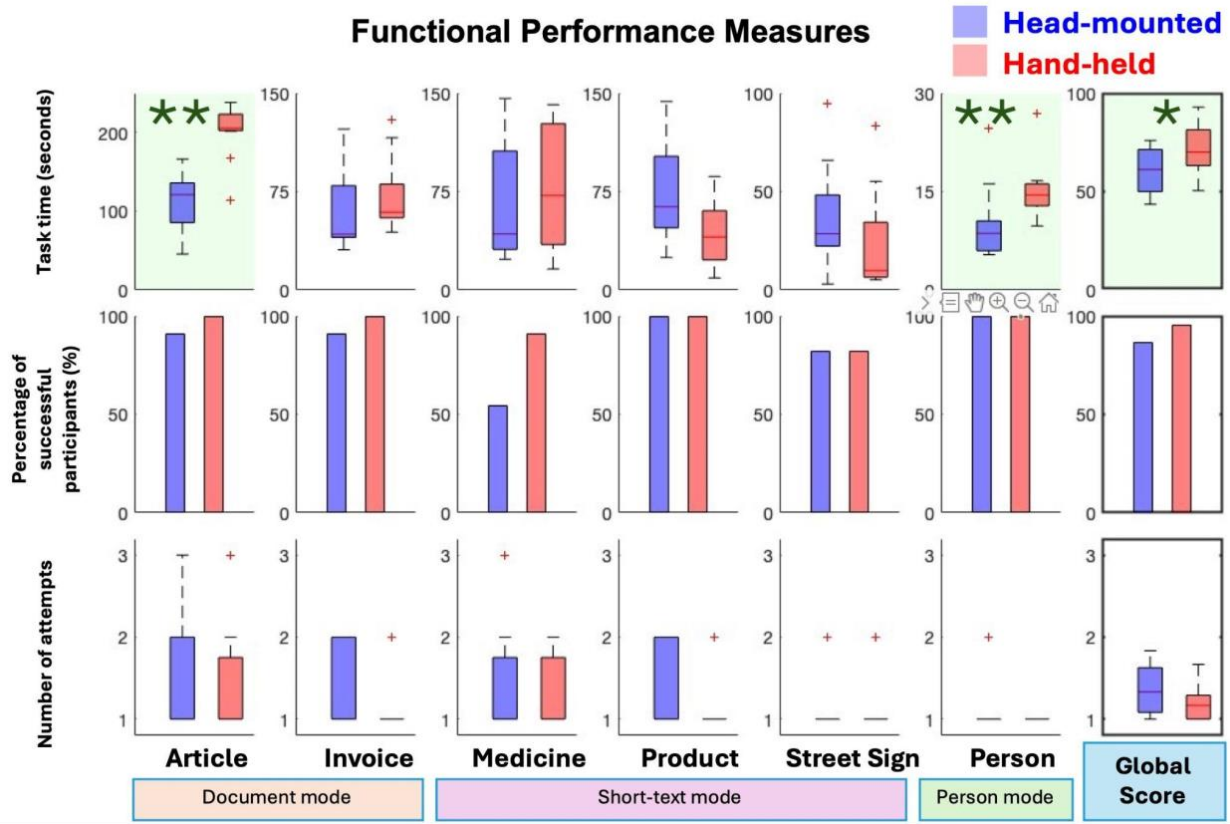

*Figure 3: Comparison of head-mounted (blue) and hand-held (red) systems for functional parameters during the final attempt. (A) Total Task Time, (B) Percentage of Successful Participants, (C) Number of Attempts. Statistically significant differences are indicated by green shading. A single asterisk (*) denotes BH-corrected $p < 0.05$, and a double asterisk (**) denotes BH-corrected $p < 0.01$.*

### Biomechanical Measures

The table Table 3 below provides the summary and statistics of biomechanical measures. More detailed results for each parameter are provided in the following sections.

*Table 3: **Biomechanical Parameters Across Tasks and Embodiments for Final Attempt**. The table summarizes the average of biomechanical metrics across all*

participants for each task, comparing the head-mounted (HM) and hand-held (HH) systems. Metrics include joint ROM and angular path length for head, trunk and bilateral hands, and working volume and movement smoothness for head and bilateral hands. The third row in each body segment reports Benjamini-Hochberg (BH) corrected p-values from Wilcoxon signed-rank tests. Shaded cells indicate statistically significant differences ($p < 0.05$).

| Body Segment | Embodiment Type | Article | Invoice | Medicine | Product | Street Sign | Person | Global Score |
|---|---|---|---|---|---|---|---|---|
| \multicolumn{9}{c}{Joint ROM (°)} |
| Head | HM | 37.97 | 34.97 | 33.77 | 47.78 | 31.13 | 13.97 | 33.27 |
| | HH | 42.94 | 36.68 | 33.74 | 41.56 | 27.12 | 13.05 | 32.52 |
| | p | 0.524 | 0.966 | 0.966 | 0.524 | 0.697 | 0.697 | 0.465 |
| Trunk | HM | 3.90 | 2.83 | 3.56 | 4.46 | 1.52 | 1.56 | 2.97 |
| | HH | 6.34 | 4.77 | 4.07 | 4.00 | 2.79 | 1.94 | 3.98 |
| | p | 0.006 | 0.006 | 0.577 | 0.577 | 0.125 | 0.125 | 0.008 |
| Right Hand | HM | 88.58 | 80.67 | 71.34 | 70.57 | 24.82 | 7.95 | 57.32 |
| | HH | 88.60 | 80.79 | 84.15 | 71.16 | 56.39 | 27.27 | 68.06 |
| | p | 0.840 | 0.779 | 0.480 | 0.966 | 0.161 | 0.059 | 0.090 |
| Left Hand | HM | 62.36 | 57.64 | 63.79 | 91.66 | 14.74 | 11.99 | 50.36 |
| | HH | 82.81 | 70.12 | 58.61 | 65.71 | 41.86 | 29.07 | 58.03 |
| | p | 0.221 | 0.520 | 0.334 | 0.027 | 0.027 | 0.012 | 0.084 |
| \multicolumn{9}{c}{Angular Path Length (°)} |
| Head | HM | 146.23 | 122.09 | 145.77 | 313.16 | 131.20 | 27.71 | 147.69 |
| | HH | 300.95 | 149.77 | 140.42 | 185.86 | 90.33 | 32.20 | 149.92 |
| | p | 0.006 | 0.412 | 0.765 | 0.097 | 0.480 | 0.496 | 0.966 |
| Trunk | HM | 22.81 | 12.92 | 15.66 | 29.51 | 7.46 | 4.41 | 15.46 |
| | HH | 44.17 | 18.17 | 17.76 | 17.60 | 7.61 | 6.01 | 18.55 |
|  | p | 0.012 | 0.262 | 0.700 | 0.166 | 0.700 | 0.097 | 0.074 |
| Right Hand | HM | 339.25 | 356.25 | 328.05 | 666.53 | 78.87 | 14.73 | 297.28 |
|  | HH | 659.20 | 579.40 | 232.24 | 441.68 | 160.18 | 47.16 | 353.31 |
|  | p | 0.056 | 0.496 | 0.765 | 0.480 | 0.480 | 0.006 | 0.641 |
| Left Hand | HM | 208.75 | 190.16 | 207.25 | 841.71 | 66.96 | 29.17 | 257.33 |
|  | HH | 496.29 | 278.58 | 188.58 | 425.76 | 95.32 | 49.88 | 255.73 |
|  | p | 0.146 | 0.558 | 0.831 | 0.221 | 0.221 | 0.221 | 0.966 |
| Working Volume ($cm^3$) | | | | | | | | |
| Right Hand | HM | 15568 | 17881 | 12718 | 49152 | 3149 | 1187 | 16609 |
|  | HH | 36664 | 30066 | 9639 | 44104 | 25101 | 29126 | 29117 |
|  | p | 0.041 | 0.048 | 0.966 | 0.558 | 0.041 | 0.048 | 0.015 |
| Left Hand | HM | 12549 | 6997 | 12141 | 60302 | 396 | 5725 | 16352 |
|  | HH | 37765 | 21671 | 10139 | 42843 | 16907 | 27263 | 26098 |
|  | p | 0.056 | 0.262 | 0.334 | 0.365 | 0.018 | 0.135 | 0.042 |
| Head | HM | 586 | 313 | 321 | 347 | 21 | 31 | 269 |
|  | HH | 5793 | 3080 | 423 | 340 | 15 | 18 | 1611 |
|  | p | 0.029 | 0.029 | 0.697 | 1.000 | 0.840 | 0.107 | 0.006 |
| Movement Smoothness (dimensionless) | | | | | | | | |
| Right Hand | HM | -17.71 | -15.61 | -15.82 | -16.67 | -14.32 | -10.81 | -15.16 |
|  | HH | -20.47 | -16.87 | -16.48 | -14.52 | -11.33 | -11.04 | -15.12 |
|  | p | 0.006 | 0.309 | 0.700 | 0.135 | 0.135 | 0.558 | 0.765 |
| Left Hand | HM | -17.79 | -15.92 | -15.98 | -16.48 | -14.46 | -10.14 | -15.13 |
|  | HH | -19.98 | -16.48 | -16.00 | -14.51 | -11.29 | -11.44 | -14.95 |
|  | p | 0.006 | 0.558 | 0.765 | 0.152 | 0.152 | 0.152 | 0.765 |
| Head | HM | -18.53 | -16.24 | -16.43 | -17.48 | -14.61 | -11.01 | -15.72 |

| | HH | -20.14 | -16.57 | -16.61 | -15.18 | -12.38 | -12.90 | -15.63 |
| | p | 0.006 | 0.966 | 0.966 | 0.064 | 0.221 | 0.009 | 0.765 |

***Joint ROM:*** Joint ROM varied by body segment and task across the two embodiments (Figure 4, Table 3). The trunk exhibited significantly greater ROM with the hand-held system during the Article and Invoice tasks, and this trend was also reflected in the global scores, which showed a 35% increase in trunk movement compared to the head-mounted system. For the left hand, ROM was significantly higher with the head-mounted system in the Product task, but higher with the hand-held system in the Person and Street Sign tasks. No significant differences were found for right-hand or head ROM, either at the task level or in global scores.

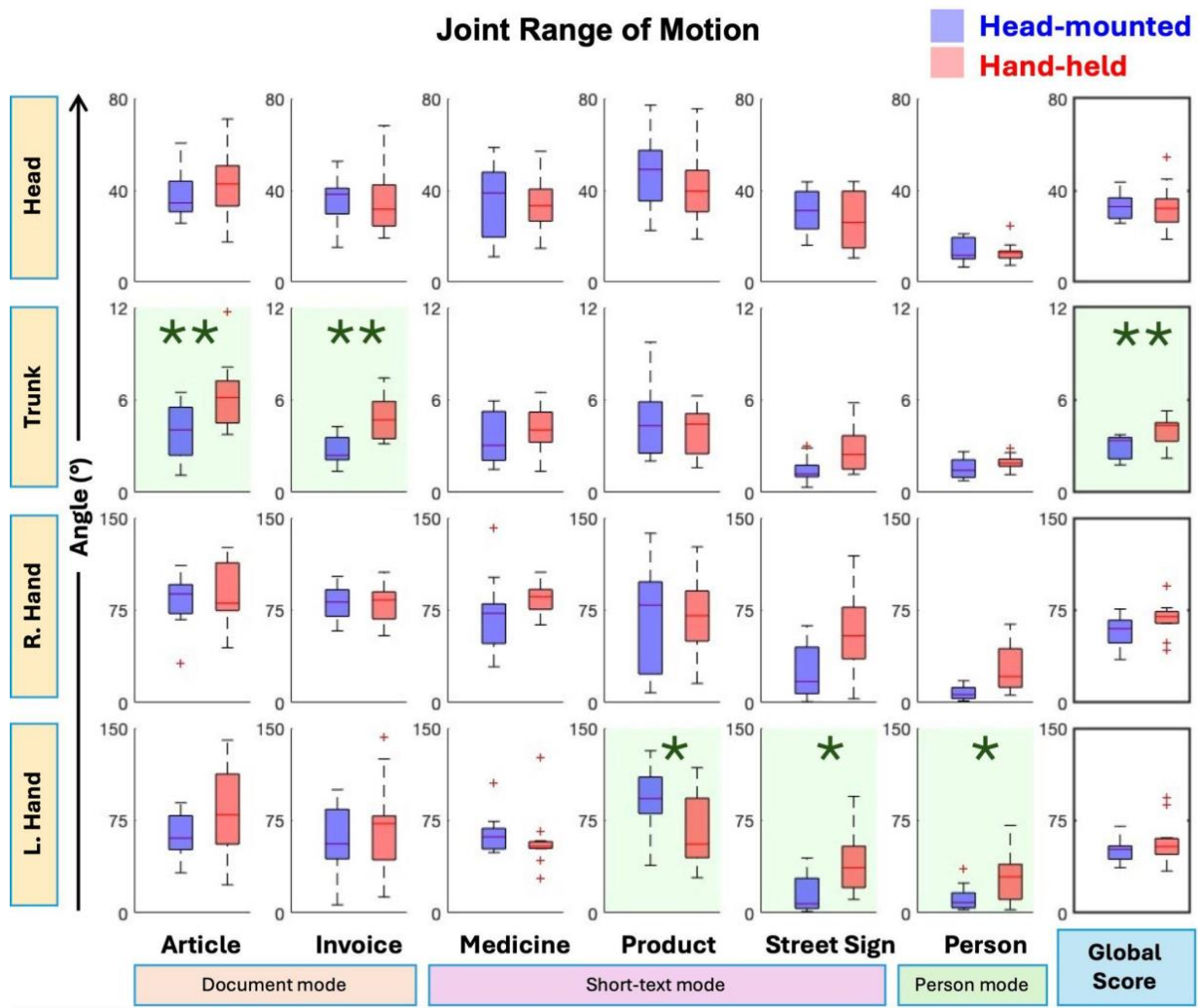

*Figure 4: Comparison of head-mounted (blue) and hand-held (red) embodiments for Joint ROM during final attempt. A higher value of Joint ROM indicates higher angular displacement of the joint segment, reflecting increased movement range. Statistically significant differences are indicated by green shading. A single asterisk (\*) denotes*

BH-corrected p < 0.05, and a double asterisk (**) denotes BH-corrected p < 0.01.

**Angular Path Length:** Angular path length varied by body segment and task across the two embodiments (Figure 5, Table 3). The hand-held system showed significantly greater angular path length in the head and trunk during the Article task, and in the right hand during the Person task. Although no significant differences were found for the left hand, descriptive trends indicated consistently larger angular paths with the hand-held system across most tasks. Global scores did not differ significantly between systems for any body segment.

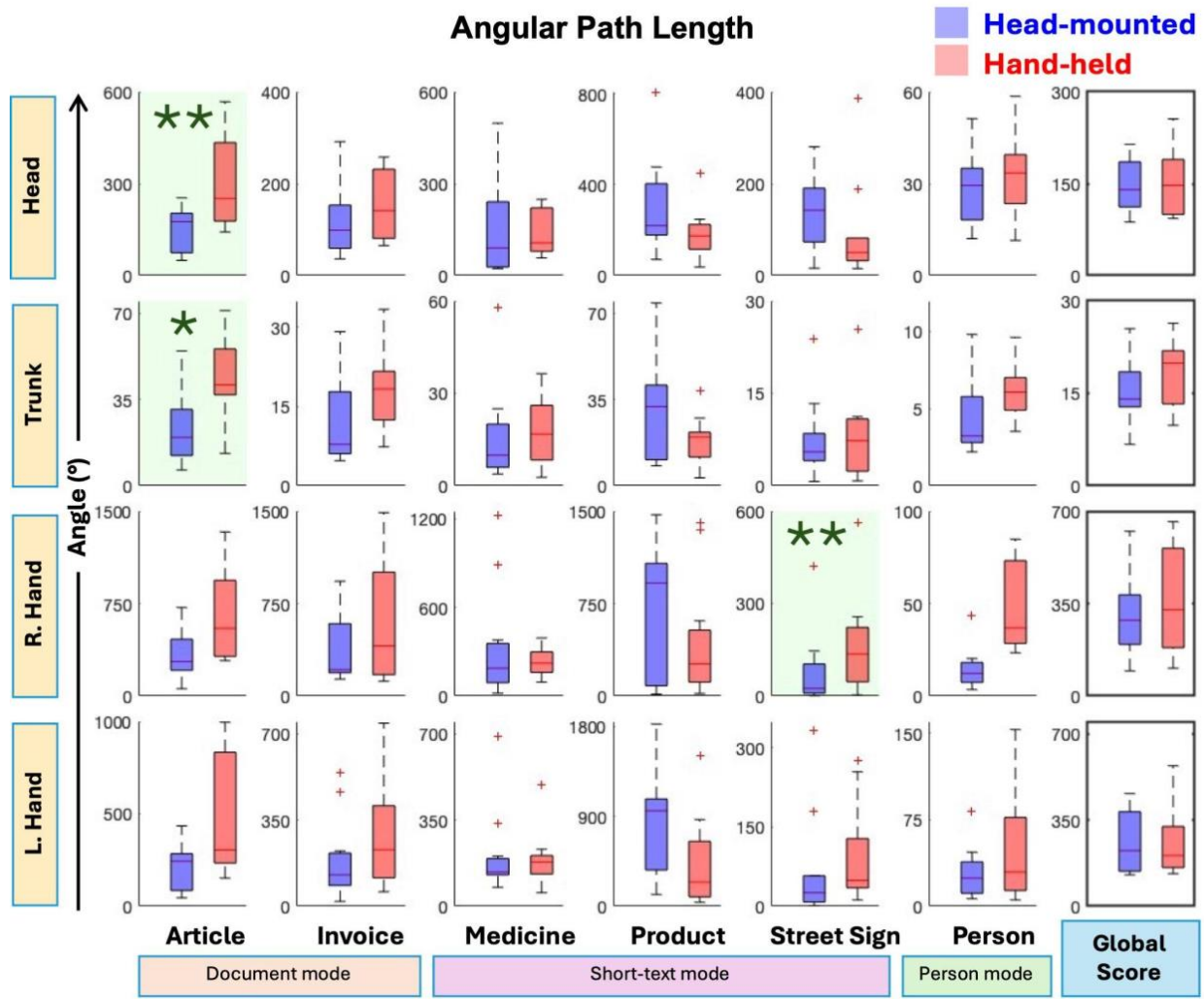

Figure 5: Comparison of head-mounted (blue) and hand-held (red) embodiments for Angular path length during final attempt. A higher value of Angular Path length indicates higher angular motion of the joint segment, reflecting increased movement quantity. Statistically significant differences are indicated by green shading. A single asterisk (*) denotes BH-corrected p < 0.05, and a double asterisk (**) denotes BH-corrected p < 0.01.

***Working Volume:*** Working volume was consistently greater with the hand-held system across all body segments studied (Figure 6, Table 3). Significant differences were observed in the right hand during the Article, Invoice, Person, and Street Sign tasks; in the left hand during the Street Sign task; and in the head during the Article and Invoice tasks with the hand-held system exhibiting higher working volumes. Global scores reinforced this pattern, with the hand-held system showing a 75% increase in right-hand working volume, 60% increase in the left hand, and a nearly 500% increase in the head segment. These findings suggest that the hand-held system required substantially more spatial movement, particularly during tasks involving document handling and environmental scanning.

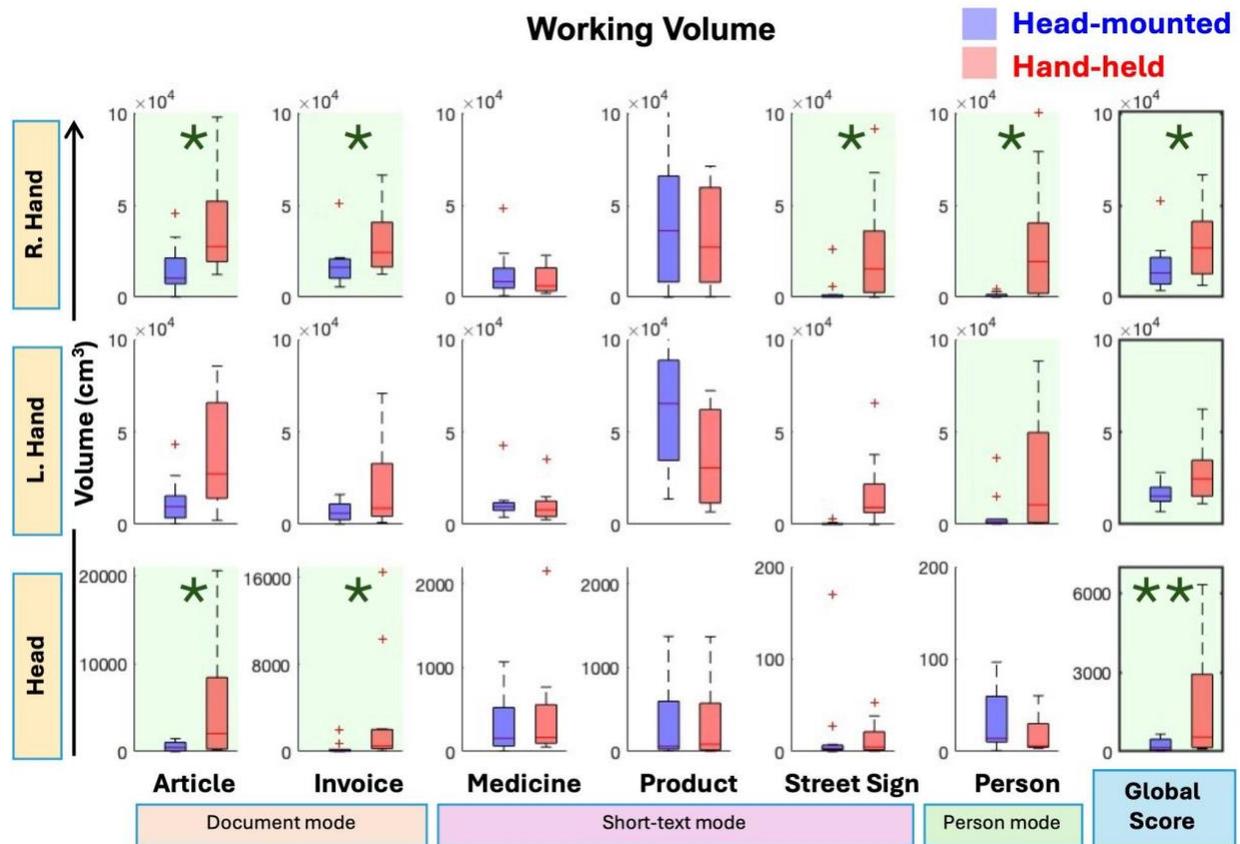

*Figure 6: Comparison of head-mounted (blue) and hand-held (red) systems for Working Volume during final attempt. A higher value of Working Volume indicates higher spatial coverage during task execution. Statistically significant differences are indicated by green shading. A single asterisk (\*) denotes BH-corrected $p < 0.05$, and a double asterisk (\*\*) denotes BH-corrected $p < 0.01$.*

***Movement Smoothness:*** Movement smoothness measured using log dimensionless jerk - velocity (LDLJ-V), varied by task and body segment across the two embodiments (Figure 7, Table 3), with higher LDLJ-V values indicating smoother motion. The head-mounted system produced significantly smoother movements in the right hand, left hand, and head during the Article task, and in the head segment during the Person task.

While no significant differences were observed in the global smoothness scores across any of the body segments, the head-mounted system consistently showed higher average smoothness values across the board.

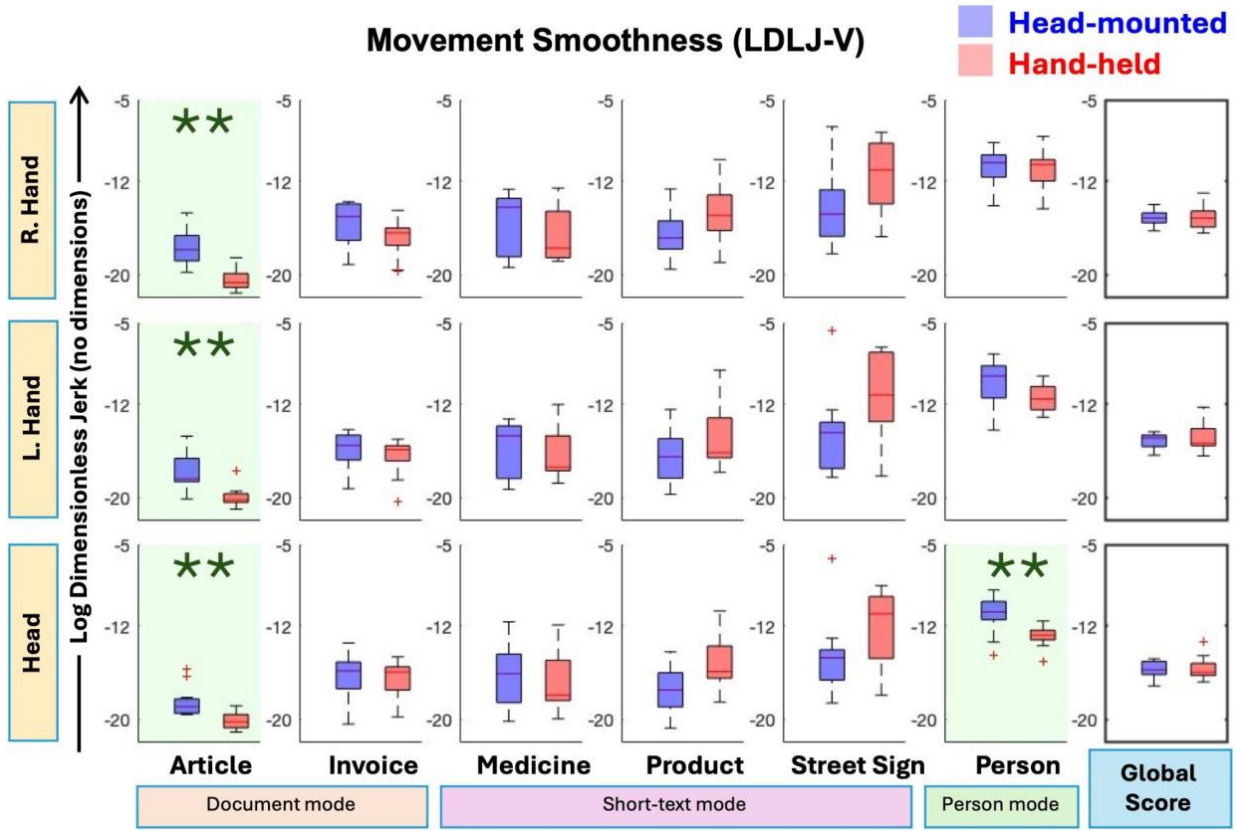

Figure 7: Comparison of head-mounted (blue) and hand-held (red) systems for movement smoothness (LDLJ-V) during final attempt. A higher value of LDLJ-V indicates higher movement smoothness. Statistically significant differences are indicated by green shading. A single asterisk (*) denotes BH-corrected $p < 0.05$, and a double asterisk (**) denotes BH-corrected $p < 0.01$.

**Secondary Analysis - Summative Performance**

The summative analysis considered cumulative task performance across all attempts, revealing additional trends not captured by the final-attempt analysis. The summaries are provided in Tables 4 and 5.

Table 4: **Task Time Across Tasks and Embodiments for Summative Analysis**. The table summarizes the average task time across all participants for each task, comparing the head-mounted (HM) and hand-held (HH) systems. The third-row reports Benjamini-

Hochberg (BH) corrected p-values from Wilcoxon signed-rank tests. Shaded cells indicate statistically significant differences (p < 0.05).

| Embodiment Type | Article | Invoice | Medicine | Product | Street Sign | Person | Global Score |
|---|---|---|---|---|---|---|---|
| Task Time (seconds) | | | | | | | |
| HM | 176.0 | 79.1 | 114.8 | 108.3 | 44.1 | 10.7 | 88.9 |
| HH | 243.0 | 86.0 | 90.7 | 62.0 | 31.3 | 15.1 | 88.0 |
| p | 0.009 | 0.966 | 0.558 | 0.010 | 0.152 | 0.009 | 0.831 |

**Task Time:** Significant differences in task time were observed for the Article, Product, and Person tasks. As with the final attempt analysis, the head-mounted system was faster for Article and Person tasks. However, there was an added statistical significance in the Product task with the hand-held system being faster. No significant differences were found in the global task time.

Table 5: **Biomechanical Parameters Across Tasks and Embodiments for Summative Analysis**. *The table summarizes the average of biomechanical metrics across all participants for each task, comparing the head-mounted (HM) and hand-held (HH) systems. Metrics include joint ROM and angular path length for head, trunk and bilateral hands, and working volume and movement smoothness for head and bilateral hands. The third row in each body segment reports Benjamini-Hochberg (BH) corrected p-values from Wilcoxon signed-rank tests. Shaded cells indicate statistically significant differences (p < 0.05).*

| Body Segment | Embodiment Type | Article | Invoice | Medicine | Product | Street Sign | Person | Global Score |
|---|---|---|---|---|---|---|---|---|
| Joint ROM (°) | | | | | | | | |
| Head | HM | 40.29 | 37.52 | 38.20 | 50.26 | 31.68 | 14.14 | 35.35 |
| | HH | 44.11 | 37.94 | 34.97 | 41.64 | 29.22 | 13.05 | 33.49 |
| | p | 0.693 | 0.966 | 0.693 | 0.146 | 0.693 | 0.693 | 0.275 |
| Trunk | HM | 4.18 | 3.31 | 3.85 | 5.20 | 1.61 | 1.61 | 3.29 |
| | HH | 6.61 | 4.99 | 4.10 | 4.06 | 2.79 | 1.94 | 4.08 |
|  | p | 0.009 | 0.009 | 0.966 | 0.177 | 0.125 | 0.125 | 0.098 |
| Right Hand | HM | 95.79 | 81.91 | 76.91 | 73.61 | 25.76 | 8.30 | 60.38 |
|  | HH | 94.22 | 84.98 | 85.73 | 73.90 | 59.10 | 27.27 | 70.87 |
|  | p | 0.918 | 0.620 | 0.557 | 0.966 | 0.161 | 0.059 | 0.107 |
| Left Hand | HM | 77.17 | 58.79 | 67.39 | 97.49 | 14.79 | 12.26 | 54.65 |
|  | HH | 85.21 | 73.61 | 60.74 | 69.71 | 44.07 | 29.07 | 60.40 |
|  | p | 0.898 | 0.496 | 0.360 | 0.049 | 0.041 | 0.012 | 0.320 |
| **Angular Path Length (°)** | | | | | | | | |
| Head | HM | 237.05 | 169.01 | 251.10 | 379.21 | 153.81 | 29.82 | 203.33 |
|  | HH | 399.10 | 215.42 | 176.36 | 223.40 | 107.85 | 32.20 | 192.39 |
|  | p | 0.097 | 0.480 | 0.496 | 0.006 | 0.480 | 0.700 | 0.577 |
| Trunk | HM | 32.79 | 16.78 | 24.55 | 38.67 | 8.52 | 4.95 | 21.04 |
|  | HH | 56.91 | 24.34 | 20.43 | 21.62 | 8.61 | 6.01 | 22.99 |
|  | p | 0.018 | 0.262 | 0.558 | 0.029 | 0.700 | 0.262 | 0.577 |
| Right Hand | HM | 504.03 | 432.21 | 495.73 | 949.53 | 81.10 | 17.36 | 413.33 |
|  | HH | 930.58 | 697.85 | 276.30 | 470.89 | 191.80 | 47.16 | 435.76 |
|  | p | 0.073 | 0.413 | 0.288 | 0.107 | 0.288 | 0.012 | 0.577 |
| Left Hand | HM | 345.34 | 257.69 | 359.54 | 1127.90 | 67.90 | 33.37 | 365.29 |
|  | HH | 659.83 | 375.95 | 239.82 | 573.98 | 110.61 | 49.88 | 335.01 |
|  | p | 0.221 | 0.765 | 0.334 | 0.221 | 0.221 | 0.221 | 0.577 |
| **Working Volume (cm³)** | | | | | | | | |
| Right Hand | HM | 21978 | 20318 | 19147 | 69258 | 3161 | 1421 | 22547 |
|  | HH | 46325 | 34529 | 11201 | 44706 | 28033 | 29126 | 32320 |
|  | p | 0.048 | 0.041 | 0.278 | 0.278 | 0.041 | 0.048 | 0.032 |
| Left Hand | HM | 16052 | 7814 | 19615 | 72852 | 398 | 5888 | 20436 |
|  | HH | 51090 | 26829 | 12940 | 53087 | 17503 | 27263 | 31452 |
|  | p | 0.041 | 0.177 | 0.125 | 0.32 | 0.018 | 0.125 | 0.010 |
| Head | HM | 715 | 403 | 416 | 500 | 22 | 44 | 350 |
|  | HH | 8696 | 3773 | 459 | 346 | 16 | 18 | 2218 |
|  | p | 0.021 | 0.021 | 0.831 | 0.548 | 0.831 | 0.107 | 0.010 |
| **Movement Smoothness (dimensionless)** | | | | | | | | |
| Right Hand | HM | -19.29 | -16.65 | -17.81 | -17.90 | -14.52 | -11.09 | -16.21 |
|  | HH | -21.09 | -17.32 | -17.20 | -15.09 | -11.67 | -11.04 | -15.57 |
|  | p | 0.021 | 0.697 | 0.765 | 0.021 | 0.166 | 0.966 | 0.221 |
| Left Hand | HM | -19.49 | -16.89 | -17.98 | -17.74 | -14.73 | -10.41 | -16.21 |
|  | HH | -20.64 | -17.01 | -16.75 | -15.02 | -11.68 | -11.44 | -15.42 |
|  | p | 0.073 | 0.966 | 0.438 | 0.073 | 0.135 | 0.185 | 0.320 |
| Head | HM | -19.93 | -17.20 | -18.40 | -18.56 | -15.02 | -11.28 | -16.73 |
|  | HH | -20.71 | -16.94 | -17.42 | -15.70 | -12.73 | -12.90 | -16.07 |
|  | p | 0.027 | 0.700 | 0.700 | 0.015 | 0.185 | 0.015 | 0.221 |

*Joint ROM:* Trunk and left-hand ROM showed task-specific differences between embodiments. The hand-held system required significantly greater trunk ROM in the Article and Invoice tasks, and greater left-hand ROM in the Person and Street Sign tasks. In contrast, during the Product task, the head-mounted system showed greater left-hand ROM. No significant differences were observed in the global ROM scores for any body segment.

*Angular Path Length:* Angular path length showed varied task-specific differences across body segments. The hand-held system produced greater angular path length in the trunk segment for the Article task, and in the right-hand segment for the Person task. The head-mounted system showed greater trunk and head angular path length in the Product task. No significant differences were observed in the global angular path length scores for any of the body segments.

***Working Volume:*** The hand-held system consistently produced greater working volume across multiple body segments and tasks. Significant differences were observed in the right hand during the Article, Invoice, Person, and Street Sign tasks, and in the left hand during the Article and Street Sign tasks. The head segment also showed significantly greater working volume with the hand-held system in the Article and Invoice tasks. Global scores reflected this trend, with the hand-held system showing approximately 40% greater working volume in the right hand, 50% in the left hand, and a 530% increase in the head segment, indicating a broader spatial range of movement when using the hand-held embodiment.

***Movement Smoothness (LDLJ-V):*** Movement smoothness showed task- and body segment-specific differences between embodiments. The head-mounted system produced significantly smoother right-hand movements during the Article task, while the hand-held system was smoother in the Product task. In the head segment, the head-mounted system showed smoother motion in the Article and Person tasks, whereas the hand-held system showed smoother head movement in the Product task. No significant differences were observed for the left hand or in global smoothness scores across any of the body segments.

## DISCUSSION

Across tasks and analyses, a consistent trend emerged in our findings across several task and body-segment combinations: the head-mounted system generally required less upper-body motion and had shorter task completion times, while the hand-held system exhibited, on average, higher success rates, requiring fewer attempts. These patterns held across multiple biomechanical metrics, most notably in working volume, but also in joint range of motion and angular path length, suggesting that although both systems are functionally viable, they differ meaningfully in the biomechanical cost associated with use. Notably, smoother movements were also more frequently observed with the head-mounted system in several task and body-segment combinations, further reinforcing the idea that embodiment plays a critical role in shaping movement strategies and physical efficiency.

These results highlight a perspective often overlooked in the evaluation of assistive technology: not just whether users can complete a task, but what kind of physical demand is placed on them in the process. This is particularly relevant for pBLV, who are already at increased risk for musculoskeletal disorders and may experience cumulative strain from repetitive or compensatory movements (Kolli et al. 2022; Julie Mount, Gitlin, and Howard 1997; McCall et al. 2021; Gitlin et al. 1997). While functional outcomes have provided valuable insights in prior research and have generally been the central focus, our findings emphasize that incorporating objective biomechanical indicators can offer a more comprehensive understanding of device usability and physical demand.

Taken together, the results suggest that embodiment form, the physical configuration and interaction style of the system, meaningfully influences both performance and effort. This motivates a more nuanced understanding of what it means for an assistive technology to be "usable," pushing beyond binary measures of success, to consider the physical sustainability of interaction over time. In the following sections, we expand on these findings by examining how embodiment shaped movement patterns, how intuitive performance unfolded across attempts, and how different task types exposed distinct strengths and limitations of each embodiment.

**Embodiment Shapes Body Motion**

One of the most consistent trends observed across tasks was the difference in upper-body movement quantity between the two embodiments. The hand-held system generally required a greater joint range of motion, larger angular path lengths, and more expansive working volumes across different body segments. These findings point toward a less economical movement profile, suggesting that the physical effort required to complete a task with the hand-held system is higher compared to the head-mounted alternative.

This distinction in movement quantity is not merely a matter of efficiency; it has important implications for fatigue, overuse, and long-term accessibility (Wong, Cluff, and Kuo 2021; Fang et al. 2007; Zhang et al. 2019). For pBLV, the risk of musculoskeletal strain is already elevated, with higher reported rates of overuse injuries such as carpal tunnel syndrome and chronic upper-limb discomfort —often attributed to prolonged use of white canes or the physical demands of guide-dog handling (Julie Mount, Gitlin, and Howard 1997; McCall et al. 2021; Gitlin et al. 1997). Introducing assistive technologies that compound these physical demands, even subtly, may unintentionally exacerbate pre-existing vulnerabilities. In this context, the economy of motion afforded by the head-mounted system becomes a critical ergonomic advantage, particularly for users with limited muscular endurance, joint mobility, or pre-existing strain.

Among the biomechanical variables assessed, working volume consistently showed significant differences between the two systems across tasks. A larger working volume implies not only more movement but likely a greater degree of muscular engagement to execute the same action, raising concerns about potential fatigue or physical strain, particularly with repeated use. Such movement inefficiency may pose a barrier to adoption or sustained use for populations already susceptible to strain. Similar patterns were seen in the angular kinematics and joint range of motion data, which also measured how much the body moved during each task. While working volume showed the strongest effects, all three metrics pointed to a higher-movement demand with the hand-held system. Global scores support this finding: the hand-held system led to 35% greater trunk ROM, and larger working volumes in the right hand (75%), left hand (60%), and head (500%) compared to the head-mounted system. This increased movement can be attributed to the physical strategies participants used to position the smartphone camera effectively. Although the head-mounted system involved some body movement, the hand-held system required greater use of multiple joints and body segments to position the camera and capture the object effectively. For example, during the article-reading task, several participants stood up to fit the entire newspaper article into the frame, increasing motion across all body segments. Participants with residual vision may have further adjusted their posture to visually confirm alignment, contributing to elevated head and trunk movement.

In addition to movement quantity, we also evaluated movement smoothness, which serves as a proxy for movement quality. While metrics like joint ROM and working volume capture how much motion was required, smoothness reflects how fluid or erratic that motion was. In our analysis, smoother movements were more frequently observed with the head-mounted system during tasks like Article reading and Person detection. These smoother profiles may indicate fewer within-trial corrections, such as readjusting the device or re-aiming the camera to capture the desired input. In contrast, jerkier motion patterns in the hand-held condition may reflect micro-adjustments as users attempt to stabilize the phone or find the correct positioning through repeated small shifts. Over time, such jerkier movements can impose greater mechanical strain on joints and soft tissues, potentially accelerating wear-and-tear processes and contributing to musculoskeletal conditions.

**Performance Across Attempts**

While the head-mounted system consistently showed advantages in terms of biomechanical efficiency and total task times, participants generally required more trials and had lower success rates compared to the hand-held system. This pattern was observed across several tasks, suggesting that the head-mounted configuration may be less intuitive to use, especially for users unfamiliar with its form factor or camera positioning. Rather than contradicting the system's ergonomic advantages, this finding highlights the need to consider usability and learning curves as distinct yet complementary aspects of assistive-technology evaluation.

We analyzed performance using two approaches: one based on the final successful attempt and another that summed performance across all attempts. These two views revealed different aspects of system use. The final attempt analysis often highlighted the biomechanical efficiency of the head-mounted system, showing reduced movement across several task and body-segment combinations. It also showed shorter task completion times in the Article reading and Person detection tasks for the head-mounted system. In contrast, the summative analysis captured the total physical effort required to complete each task and revealed that, while the head-mounted system remained more efficient in most cases, significant differences favoring the hand-held system emerged in the Product identification task for completion time and several biomechanical metrics. As shown in tables 2-5, most of the statistically significant findings were consistent across both types of analysis, which reinforces the robustness of the observed effects. Looking only at the final outcome may cause one to miss the effort it took to get there. Usability is not just about completing a task, but also about how easy and physically demanding it was. This aligns with human-computer interaction and accessible-design work that promotes a more holistic view of usability, including learnability and physical strain alongside success rates and task time.

We also observed that user strategies differed across embodiments, further influencing performance. Some participants using the head-mounted system preferred to move their head to align the camera, while others kept their head still and moved the object into the camera's field of view. These variations, driven by individual differences, task demands and real-time feedback from the Seeing AI app, suggest that embodiment not only affects biomechanics but also shapes interaction style. Importantly, this underscores the need for targeted training programs that help users learn effective strategies specific to different device configurations. Without such training, intuitive usability may remain low even for systems with demonstrable ergonomic benefits.

**The Effect of Task Type**

Performance patterns differed across tasks, highlighting how task demands interact with system embodiment. In tasks like Article, Invoice, Person, and Street Sign, the head-mounted system consistently showed biomechanical advantages across several task and body-segment combinations. These findings suggest that for document-style or scanning tasks, the head-mounted system may offer a more efficient user experience, likely due to its clear field of view, stable central mounting and hands-free operation, which reduce the need for frequent repositioning or manual alignment. While both systems require users to solve a spatial puzzle of sorts to align the camera with the target object, the hand-held system may pose additional challenges because its field of view is not naturally aligned with the user's head. In comparison, the head-mounted system may offer a more intuitive alignment by more closely approximating the user's natural field of view, which could help reduce some of the cognitive and physical effort involved in spatial coordination.

However, this advantage did not hold across all task types. In the Product and Medicine identification tasks, which involved short text recognition on curved surfaces, there were few significant differences when analyzing final attempts. In the summative analysis, the hand-held system showed better biomechanical performance in the Product identification task across several measures. These tasks likely challenged the head-mounted system due to technical limitations rather than embodiment alone. The head-mounted camera's relatively lower image quality, combined with the difficulty of maintaining a stable focal distance on cylindrical objects, likely affected Optical Character Recognition (OCR) accuracy and forced users to make more corrective movements. OCR algorithms are also less reliable when text curves around surfaces, creating further inconsistencies in recognition. These issues may explain the lack of significant results in the Medicine identification task and the mixed performance outcomes in the Product identification task across analyses.

Task performance may also have been influenced by the Seeing AI app's operating mode. Both the Product and Medicine identification tasks used Short-Text mode, for which the hand-held system, offering better image resolution and more flexible manual framing, was likely more effective. In contrast, document-style tasks benefit from the head-mounted system's hands-free alignment and broader field of view. These findings suggest that no single embodiment is ideal for all tasks. Future assistive technologies should support task-specific adaptation, either through multimodal systems or tailored software and hardware configurations. Tasks involving dynamic navigation (e.g., Street Sign) or large document reading (e.g., Article) may benefit from head-mounted solutions, while close-range, detail-oriented tasks with curved or occluded surfaces may still be better served by a smartphone-based hand-held interface with superior camera quality.

**Study Limitations and Considerations**

Several limitations should be considered when interpreting these findings. The study involved a relatively small sample size and short-term exposure to each device, which may not fully reflect long-term user behavior. In addition, the camera quality of the smartphone used in the hand-held condition was substantially better than the head-mounted camera, likely influencing OCR performance—particularly in tasks requiring focus on curved surfaces. These hardware differences may have contributed to the higher success rates seen with the hand-held embodiment and should be controlled or normalized in future comparisons. Additionally, participants may have entered the study with greater familiarity and prior experience using smartphones for a range of tasks, potentially biasing performance in favor of the hand-held system. Future studies should consider assessing and controlling for such differences in baseline experience across device modalities.

**Implications for Design, Training, and Research**

This study highlights the need for assistive-technology design to move beyond functional success alone and consider the biomechanical demands of use. Functional outcomes on their own may provide an incomplete or even misleading picture of usability; the head-mounted system, for example, showed fewer successful trials yet delivered clear biomechanical advantages in scanning-style tasks. These nuances underscore the importance of developing data-driven, multi-dimensional evaluation frameworks that balance success rates with measures such as physical load, learnability, and adaptability. While the head-mounted system showed clear benefits in reducing upper-body movement, it often required more attempts for task completion—pointing to issues of intuitiveness, familiarity, and potentially device image quality limitations, not necessarily limitations in the form factor itself. Many participants were more familiar with smartphone use than head-mounted interfaces, which may have contributed to the performance gap. This underscores the importance of targeted training protocols, both for head-mounted systems like ARx and for smartphone-based apps, to ensure users are equipped to use each embodiment effectively.

Researchers and designers should prioritize not only usability but also long-term ergonomic sustainability, especially for technologies intended for daily use. Considering that pBLV already face elevated risks of musculoskeletal strain due to regular use of mobility aids (e.g., white canes, guide dogs), reducing unnecessary physical burden should be seen as a preventive-design goal. Integrating biomechanical considerations into early stages of design could help reduce overuse injuries and improve user comfort over time.

The findings also support the idea that no single embodiment fits all tasks. Future systems should allow for task-adaptive interaction, either through multimodal platforms or flexible design features that accommodate different scenarios. Head-mounted systems may be best suited for tasks that involve reading or environmental scanning, while hand-held devices may be more effective in short-range, detail-oriented tasks that benefit from precise framing and higher-resolution imaging. The ease of transitioning from non-use to active use may differ between embodiments, with wearable systems enabling quicker activation and smartphones introducing micro-barriers that could hinder regular use. Physicians may advise form factor selection based on task demands. To build on this work, future research should examine the impact of structured training on both task performance and movement efficiency, particularly over extended periods of use. Differences in image quality, ease of activation, and prior user familiarity should also be considered, as they may influence performance and long-term usability. Expanding testing to include additional Seeing AI modes, such as barcode scanning, currency recognition, and scene description, would provide a more complete picture in real-world scenarios. Finally, integrating co-design approaches that involve users directly in the development process may help ensure that both hardware and software features align more closely with the needs, preferences, and abilities of pBLV.

**CONCLUSION**

This study demonstrates that the embodiment of assistive technology, whether head-mounted or hand-held, has a meaningful impact on task performance and physical effort. While both embodiments enabled users to complete tasks, the head-mounted system offered certain biomechanical advantages, including reduced upper-body movement, reduced task completion time, and smoother motion, particularly for document scanning tasks. However, the hand-held system generally required fewer attempts and achieved higher success rates, especially for tasks involving short text on curved surfaces.

These findings underscore the importance of evaluating assistive technologies through a multi-dimensional lens that includes not only functional outcomes but also biomechanical performance. Incorporating biomechanical metrics into usability assessments can inform the design of more accessible and sustainable tools for pBLV. Moving forward, task-specific adaptations, targeted-training protocols, and user-centered design approaches will be critical to improving the real-world effectiveness of these technologies.

**Declaration of use of generative AI in scientific writing**
The authors used ChatGPT (OpenAI, versions 4 and 5) to assist with two aspects of manuscript preparation. First, draft schematic illustrations of the data collection methodology (Figures 1B and 1C) were generated with AI assistance. These were non-data-driven illustrations and were subsequently edited and finalized by the authors to ensure accuracy and clarity. Second, the AI tools were used to improve readability and refine language under full human supervision. All intellectual content, interpretations, and final figure designs remain the responsibility of the authors.